%% file: main.tex
\PassOptionsToPackage{%
  datamodel=acmdatamodel,
  style=acmnumeric,
}{biblatex}
\documentclass[runningheads,final]{llncs}
\usepackage[T1]{fontenc}

\input{template/header.tex}

\begin{document}

\title{WOOTdroid: Whole-system Online On-device Tracing for Android}
\titlerunning{WOOTdroid}

\author{Simon Althaus\inst{1}\and
Nikolaos Alexopoulos\inst{2}\and
Max M\"uhlh\"auser\inst{1}\and
Christian Reuter\inst{1}\and
Ephraim Zimmer\inst{1}}
\institute{Technical University of Darmstadt, Darmstadt, Germany\\
\email{\{althaus,reuter,zimmer\}@peasec.tu-darmstadt.de, max@tk.tu-darmstadt.de} \and
Athens University of Economics and Business, Athens, Greece\\
\email{alexopoulos@aueb.gr}}

\authorrunning{Althaus et al.}

\maketitle

\input{content/00-abstract.tex}
\input{content/01-introduction.tex}
\input{content/02-background.tex}
\input{content/03-motivating_example.tex}
\input{content/04-approach.tex}
\input{content/05-evaluation_results.tex}
\input{content/07-conclusion.tex}
\input{content/xx-acknowledgements.tex}

\printbibliography

\appendix
\input{content/xy-appendix.tex}

\end{document}

%% file: template/header.tex
\usepackage{ifluatex}

\ifluatex
  \usepackage{fontspec}
  \defaultfontfeatures{Ligatures=TeX}
\else
  \usepackage[T1]{fontenc}    
  \usepackage[utf8]{inputenc} 
\fi
\usepackage[english]{babel}   
\usepackage{microtype}        
\microtypesetup{nopatch={footnote}}
\usepackage[
  autostyle,
]{csquotes}                   
\usepackage[listings=false]{scrhack}        
\usepackage[
  newcommands
]{ragged2e}                   
\PassOptionsToPackage{%
  hyphens
}{url}                        

\input{template/header_biblatex}

\usepackage{ifdraft}        
\usepackage{comment}
\usepackage{multirow}
\usepackage{booktabs}
\usepackage{makecell}
\usepackage{hyphenat}       
\usepackage[
  colorinlistoftodos,
  textwidth=\marginparwidth,
  textsize=footnotesize,
  obeyFinal,
]{todonotes}                
  \presetkeys{todonotes}{%
    fancyline,
    color=red!30}{}
\usepackage{graphicx}
\usepackage{array}
\newcolumntype{L}[1]{>{\raggedright \arraybackslash}m{\if!#1!3cm\else#1\fi}}
\newcolumntype{C}[1]{>{\centering \arraybackslash}m{\if!#1!3cm\else#1\fi}}
\newcolumntype{R}[1]{>{\raggedleft \arraybackslash}m{\if!#1!3cm\else#1\fi}}
\usepackage[inline]{enumitem}
\usepackage[most]{tcolorbox}
    \tcbuselibrary{xparse}
\usepackage{hyperref}
\usepackage[                
  capitalise,               
  nameinlink,
]{cleveref}
  \crefname{cref:appendix}{Appendix}{Appendices}
  \Crefname{cref:appendix}{Appendix}{Appendices}
  \crefname{section}{Sect.}{Sects.}
  \Crefname{section}{Sect.}{Sects.}
  \crefname{table}{Tab.}{Tabs.}
  \Crefname{table}{Tab.}{Tabs.}

\usepackage{pifont}
\usepackage{xspace}
\usepackage{circledsteps}

\input{./template/header_macros}
\input{./template/header_glossary}

%% file: template/header_biblatex.tex
\usepackage[
  style=numeric,
  backend=biber,
  giveninits=true,
  maxnames=2,
  defernumbers=true,
  backref=false,
  sortcites=true,
  sorting=none,
  block=ragged,
]{biblatex}

\usepackage[
    enable=false,
    symbolpackage=pict2e,
]{biblatex-ext-oa}
\DeclareOpenAccessFieldUrl{eprint}{\thefield{eprint}}

%% file: template/header_macros.tex
\ifoptionfinal{}{%
  \overfullrule=1cm
}

\renewcommand{\emph}[1]{\textit{#1}}

\usepackage{framed}         
\renewenvironment{leftbar}{%
    \MakeFramed{\advance \hsize -\width \FrameRestore }%
}{%
    \endMakeFramed
}

\ifoptionfinal{}{%
  \usepackage{footnote}       
    \makesavenoteenv{leftbar}
}

\ifdefined\acksname
\else
    \newcommand\acksname{Acknowledgments}

    \RequirePackage{comment}
    \specialcomment{acks}{%
        \begingroup
        \section*{\acksname}
        \phantomsection\addcontentsline{toc}{section}{\acksname}
    }{%
        \endgroup
    }

    \makeatletter
        \newcommand\grantnum[3][]{#3%
        \def\@tempa{#1}\ifx\@tempa\@empty\else\space(\url{#1})\fi}
    \makeatother
\fi

\makeatletter
\newcommand{\linebreakand}{%
  \end{@IEEEauthorhalign}
  \hfill\mbox{}\par
  \mbox{}\hfill\begin{@IEEEauthorhalign}
}
\makeatother

\newcommand{\ra}[1]{\renewcommand{\arraystretch}{#1}}
\newcommand{\point}[1]{{\noindent\bf #1:} }


%% file: template/header_glossary.tex
\usepackage[
    style=super,
    acronym,
    nomain,
    nohypertypes={acronym}
]{glossaries}
\setacronymstyle{long-sp-short}

\newacronym{OS}{OS}{operating system}
\newacronym{BCC}{BCC}{BPF Compiler Collection}
\newacronym{eBPF}{eBPF}{extended Berkeley Packet Filter}
\newacronym{LSM}{LSM}{Linux Security Module}
\newacronym{MTE}{MTE}{Arm Memory Tagging Extension}
\newacronym{syscall}{syscall}{system call}
\newacronym{adb}{adb}{Android Debug Bridge}
\newacronym{IPC}{IPC}{Inter-Process Communication}
\newacronym{RPC}{RPC}{Remote Procedure Call}
\newacronym{AIDL}{AIDL}{Android Interface Definition Language}
\newacronym{app}{app}{application}
\newacronym{API}{API}{application programming interface}
\newacronym{ABI}{ABI}{Application Binary Interface}
\newacronym{APK}{APK}{Application Package}
\newacronym{WDSys}{WDSys}{WOOTdroid-Syscall}
\newacronym{WDBind}{WDBind}{WOOTdroid-Binder}
\newacronym{PID}{PID}{Process Identifier}
\newacronym{UID}{UID}{User Identifier}
\newacronym{VMI}{VMI}{Virtual Machine Introspection}

\DeclareRobustCommand{\ebpf}[0]{\gls{eBPF}\xspace}
\newcommand{\syscall}[0]{\gls{syscall}\xspace}
\newcommand{\syscalls}[0]{\glspl{syscall}\xspace}
\DeclareRobustCommand{\WD}[0]{WOOTdroid\xspace}
\DeclareRobustCommand{\WDSys}[0]{\gls{WDSys}\xspace}
\DeclareRobustCommand{\WDBind}[0]{\gls{WDBind}\xspace}

%% file: content/00-abstract.tex
\begin{abstract}
System auditing on Android faces two problems.
First, existing syscall tracers lose events under load, silently overwriting entries faster than a user space reader can drain them.
Second, security-relevant application behavior is mediated through Binder, Android's kernel IPC mechanism, and is therefore hidden from the syscall layer. The Binder parcels that the kernel does see carry no method names or typed arguments, a disconnect between low-level events and high-level behavior known as the semantic gap.
Existing approaches address the semantic gap either by modifying the Android platform, making them difficult to adjust to OS updates, or by instrumenting the traced application in user space, which sophisticated adversaries can evade by bypassing the instrumented framework APIs.

We present WOOTdroid, a design and prototype for on-device tracing on stock Android that addresses both problems without OS modification or application instrumentation.
WDSys, an eBPF port of eAudit-style syscall auditing, runs on current Android with at most 3.6\% Geekbench overhead and traces 33\% more syscalls than ftrace.
WDBind captures Binder parcels in the kernel and decodes them out-of-process against a framework signature table extracted via Java reflection.
We demonstrate WOOTdroid on Pixel 9 devices running Android 16 with an end-to-end case study reconstructing ten security-relevant Binder transactions.
\end{abstract}

%% file: content/01-introduction.tex
\section{Introduction}
\label{sec:intro}
Major steps have been taken to improve the security of the Android platform in the last fifteen years.
A significant strand of research has focused on malware detection~\cite{talha2015apk, mahindru2021fsdroid, mahindru2024permdroid, kshirsagar2022study}, in order to detect and remove
potentially harmful \glspl{app} from app stores before they can be installed on user devices.
Android's permission model has progressively improved~\cite{mayrhofer2021android, tuncay2024android},
offering more fine-grained configuration options, thus reducing the adverse effects of malicious or curious \glspl{app} that make it into user devices.
Furthermore, Android's application sandbox has progressively offered stricter isolation between \glspl{app}~\cite{mayrhofer2021android}, e.g., via Linux kernel security modules (SELinux) and seccomp, mitigating the impact of exploitable vulnerabilities in any given \gls{app}.
Despite this significant progress in security mechanisms, history has shown that adversaries can still successfully compromise Android devices~\cite{arntz2025android}, bypassing all existing protections.

System auditing is the primary method for mitigating such attacks, either via live detection and response, or via post-mortem analysis.
For non-mobile devices, e.g., desktop or cloud systems, recent research has shown how
logging and analyzing (via provenance tracking) \syscalls and other system-proximal functions
can facilitate powerful threat mitigation mechanisms~\cite{eAudit, ProvenanceSoK}.
While Android shares some commonalities with Linux-based non-mobile devices, the peculiarities of mobile devices and the challenges that come with adapting such an auditing approach for Android have to be addressed:
\begin{enumerate*}[label=(\arabic*)]
    \item \label{itm:resource_limit}Limited system resources -- Common desktop or cloud-based approaches, like the Linux Audit System (auditd) \cite{Auditd8AuditDaemon}, heavily rely on large storage space and high computing power. Limited availability of both aspects reduces either the overall system performance and responsiveness while monitoring, or the efficacy of the auditing and tracing approaches themselves due to a significant loss of monitoring events.
    \item \label{itm:semantic_gap}The semantic gap problem -- Android's sandboxed architecture prevents \glspl{app} from directly utilizing kernel functionalities and system resources, which would be monitorable on the kernel-level. Instead, \glspl{app} are forced to heavily utilize Android's uniquely provided \gls{IPC} mechanism called \emph{Binder} in order to employ functionalities and resources from other \glspl{app} and specialized system processes. This \gls{IPC} mechanism conceals the semantics of \gls{app} operations and keeps them unobservable from the kernel-level~\cite{eAudit}.
\end{enumerate*}

Due to these challenges, auditing specifically for mobile devices has received considerably less attention.
While approaches, such as Clearscope~\cite{ClearScope}, have been shown to be highly effective in intrusion
detection tasks, they have critical design limitations. Most importantly, they require extensive manual modifications to the OS, making them especially difficult to adjust to OS updates and other device models, resulting in them being quickly outdated and undeployable on modern devices.
Other less invasive approaches~\cite{binder-trace, BPFroid}, rely on user-space instrumentation
of \gls{API} calls to overcome the semantic gap problem,
making them detectable and evadable by reasonably capable attackers
who can completely circumvent them
(see our motivational example in Section~\ref{sec:motivational_example}).

To overcome these limitations, we present an adaptable design, called \WD, that can efficiently trace system events in Android without requiring system modifications nor user space instrumentation of \glspl{API} calls,
while enabling accurate common analysis via sufficient semantic information.
Our design consists of two modules:
\WDSys to efficiently trace \syscalls, and
\WDBind to extract relevant information from Binder transactions
and reconstruct the semantics of \gls{API} calls.
Both modules rely on \ebpf to transparently trace
events and extract relevant data fields without requiring
modifications to the operating system.
To efficiently trace \syscalls,
we adapt recent advances in efficient auditing for Linux hosts
(eAudit~\cite{eAudit}) to the Android domain.
To reconstruct the semantics of \gls{app} behaviors,
we build on Binder analysis techniques
from the malware analysis domain that rely on Virtual Machine Introspection
or modifications to the Android kernel~\cite{CopperDroid, ruggia2024unmasking},
and adapt them to work on unmodified devices via \ebpf.

\par\noindent\textbf{Contributions.} Our main contributions are the following:

\point{WDSys}
We demonstrate that eAudit-style \syscall auditing can be ported to stock Android devices. We identify and work around the Android-specific \ebpf challenges that have prevented this class of tools from running on recent Android versions, and we show that the resulting tracer imposes at most 3.6\% overhead on Geekbench while
outperforming related work on \syscall-trace completeness across the top 100 Google Play \glspl{app}.

\point{\WDBind}
We demonstrate that application behavior semantics can be recovered without modifying the OS or instrumenting the traced application. \WDBind captures Binder transaction buffers in the kernel at the \syscall entry point with \ebpf, and performs method-name resolution and typed-argument decoding
using a framework method signature table extracted ahead of time via Java reflection.
The vantage point, i.e.,\ kernel capture of the finished parcel
overcomes limitations of prior user space hooking approaches,
as we illustrate in Section~\ref{sec:motivational_example}.
We present an end-to-end case study reconstructing ten
security-relevant behaviors, including the SMS send of our motivating example.

%% file: content/02-background.tex
\section{Background and Related Work}
\subsection{Background}
\point{\Glsentrylong{eBPF}}
\label{sec:background-ebpf}
The \acrfull{eBPF} is an instruction set architecture that enables programs to run sandboxed in a privileged context, for example, the \gls{OS} kernel.
Such \gls{eBPF} programs can be attached to specific events, called hook points. When an application (or system) passes these hook points, the \gls{eBPF} program is invoked.
Tracepoints are out-of-the-box stable hook points that are guaranteed to remain in newer kernels. The more customizable kprobes and uprobes allow defining hooks in almost any kernel or application instructions, with the drawback of requiring rebuilding the affected applications or the whole kernel. Additionally, defining kprobes and uprobes is somewhat unstable, because the function names and instructions they are defined for are not guaranteed to remain consistent across application and kernel updates.
In more detail, an \gls{eBPF} program is a restricted C~program which is compiled to \gls{eBPF} bytecode. The bytecode is then attached to its hook point using the bpf \gls{syscall}.
Writing of \gls{eBPF} programs is made easier through libraries like \gls{BCC} that support a variety of small and carefully designed helper functions to ease \gls{eBPF} development while maintaining the stability of \gls{eBPF} programs.
An important aspect of \gls{eBPF} is the verifier, which validates each program before attaching it to its hook point. This verifier assures that \gls{eBPF} programs are safe to run and do not crash. In contrast to the classical programming of, for example, kernel drivers or modules, this significantly improves the stability of new kernel functionalities \cite{warrenMajorWindowsBSOD2024,NoMoreBlue}.
\medskip\point{Binder IPC}
\label{sec:background-binder}
In Android, \glspl{IPC} and \glspl{RPC} are facilitated by a mechanism called \emph{Binder} in a client-server fashion. As an example, the \texttt{LocationManager} provides access to information from the \texttt{Location\-Manager\-Service} as a service to other \glspl{app}. To advertise which activities and services an \gls{app} or manager provides and which parameters are expected when invoked, the service usually is defined and described by the \gls{AIDL}.
The exchange of messages for requesting a service and delivering the response is invoked by a specific \texttt{ioctl} \syscall and handled by the Binder driver. The communication is split up into transactions with transaction codes to indicate the method being called. Binder's tasks include locating the service process, delivering serialized (marshaled) messages from one process to the other using an \texttt{BINDER\_WRITE\_READ ioctl} \syscall, and also delivering objects like file descriptors. Furthermore, Binder also provides information about the sender identity (\gls{PID} and \gls{UID}) and death notifications, i.e., to indicate that a process has terminated. Objects are serialized and deserialized as \texttt{Parcel} objects for transport via Binder transactions~\cite{AndroidInternalsVolI,AndroidInternalsVolII}.

\medskip\point{eAudit}
\label{sec:background-eAudit}
We adapt a state-of-the-art \ebpf{}-based tracing approach from the domain of x86 Linux hosts called eAudit~\cite{eAudit} for use in Android devices.
Capturing of \syscalls in eAudit is done by the data capturing kernel component \textit{eCap}, while two user space components perform the logging (\textit{eLog}) and parsing (\textit{eParse}) tasks.
eCap attaches tracepoints to trace typical provenance \syscalls.
For \syscalls generally, the entry and exit can be traced separately with different hooks. The former provides arguments, while the latter provides the respective return value.
eCap additionally utilizes a very compact data encoding format that encodes events into one byte and reduces granularity when logging timestamps.
For capturing, per-CPU buffers (message caches) are used and once filled up, events are stored into a special \ebpf data structure called ring buffers, which allows passing them on to eLog for subsequent processing, releasing resources for eCap's capturing tasks.
This buffer design and additionally tuning of runtime parameters of eCap
makes it highly performant and minimizes the time window, which could be used by malicious processes to tamper with the data captured by eAudit.
Additionally, in eAudit \syscalls are prioritized based on their relevance for exploits and privilege escalation. 

\subsection{Related Work}
\label{sec:relwork}
\label{sec:shortcomings_existing}
Existing approaches for runtime auditing on Android devices either require extensive OS
modifications, making them quickly and severely outdated and impractical, or are based on user-space
instrumentation within the sandboxed vicinity of the very \glspl{app} and targets, that are being monitored, making them detectable and evadable.
In the remainder of this section, we provide more details on related approaches.

\point{Auditing and Provenance Tracking on real smartphones}
The few provenance tracking approaches that exist for Android, either rely on significant modifications of the Android system or have a significant impact on the overall system performance.
Auditd of the Linux Audit System has been ported to Android by Husted et al. \cite{AuditDAndroid} as AuditDAndroid, which, in addition to the need of modifying the Android system, still inherits auditd's general performance problem~\cite{eAudit} and does not capture sufficient semantic information.
ClearScope \cite{ClearScope} is a byte-level provenance tracking approach developed for DARPA by MIT researchers, incorporating and improving taint tracking approaches~\cite{TaintDroid,TaintART} to
capture very detailed execution traces. However, ClearScope heavily instruments the whole system, including modifications to the Android system, and relies on external servers to handle provenance storage.
binder-trace \cite{binder-trace} is a tool specifically tailored to capture and parse Binder events. However, it relies on instrumenting Android \glspl{app} and injecting custom function code into these \glspl{app} in user space using the dynamic instrumentation toolkit Frida, thus making it detectable and evadable.
Additionally, starting with Android 16, the availability of AOSP source code is uncertain \cite{AOSPInternalDev}, which breaks the way structures of captured Binder events are determined \cite{binderTraceBlogP2}. Indeed, while development is quite active, the last Android version for which required struct definitions are provided is Android 14.
BPFRoid~\cite{BPFroid} is one of the few \ebpf-based tracing approaches for Android. \ebpf is used to trace Android \gls{API} calls, native library calls, \syscalls and kernel functions.
However, for tracing Android \gls{API} functions, uprobes are used (userspace instrumentation), which makes it detectable and evadable by applications under investigation. Additionally, the parameters of \gls{API} calls are not traced, which reduces important information captured. It has been shown that the tracing of BPFRoid incurs significant data loss and runtime overhead, especially under load \cite{eAudit}.
Finally, btrace~\cite{btrace} is a sparsely documented tool that traces binder transactions by hooking a kprobe into \texttt{binder\_transaction}. However, it only parses the buffer to determine the interface token and method name, not considering method call parameters.

\point{Semantic Reconstruction}
CopperDroid \cite{CopperDroid} reconstructs Android \gls{app} behavior based on \syscalls. \Glspl{syscall} are recorded using \gls{VMI}, limiting the applicability of this approach to emulators only. Additionally, CopperDroid needs an off-device Oracle that receives encoded Android Binder, i.e., \gls{IPC} and \gls{RPC}, communication data and method signature to derive decoded parameters and encased semantics.
This off-device Oracle prevents the online on-device tracing and analysis of \glspl{app} while they are running. Despite its limitations, techniques employed in CopperDroid to parse Binder buffers are
used by \WD, adapted to run in \ebpf.
Nisi et al. \cite{NisiReconstruction} investigated how to derive higher-level \gls{API} utilization on Android just by tracing low-level \syscalls, addressing parts of the semantic gap problem. However, the authors discovered significant challenges and obstacles. Additionally, Android Binder events have not been considered.
SliceDroid, proposed by \textcite{alexopoulos2025slicedroid}, investigates the possibilities of reconstructing the semantics and behavior of \glspl{app} from tracing kernel events.
This work shows that only coarse-grained information can be reconstructed (e.g. which device was accessed) without considering the internals of Android's \gls{IPC}.

%% file: content/03-motivating_example.tex
\section{Motivating Example}
\label{sec:motivation_alternative}
\label{sec:motivational_example}
To make the challenges concrete and to illustrate why both of
\WD's contributions are necessary, we consider a recurring task in
Android auditing and provenance tracking: producing a faithful
record of an SMS send performed by an application on the device.
SMS has been a frequent target of malicious and privacy-invasive
behavior on Android, ranging from premium-rate toll fraud to the
exfiltration of authentication codes intercepted from the
notification channel~\cite{groupib2025wonderland,ruggia2025darkside}.
A useful audit record for such behavior should identify which
application sent a message, to which destination address, and with
which content, in a form suitable for subsequent analysis. We use
this task to expose two shortcomings of existing tracing approaches
that \WD addresses, and we return to it as a case study in
\cref{sec:eval}.

\begin{figure}[h]
    \centering
    \includegraphics[width=0.8\linewidth]{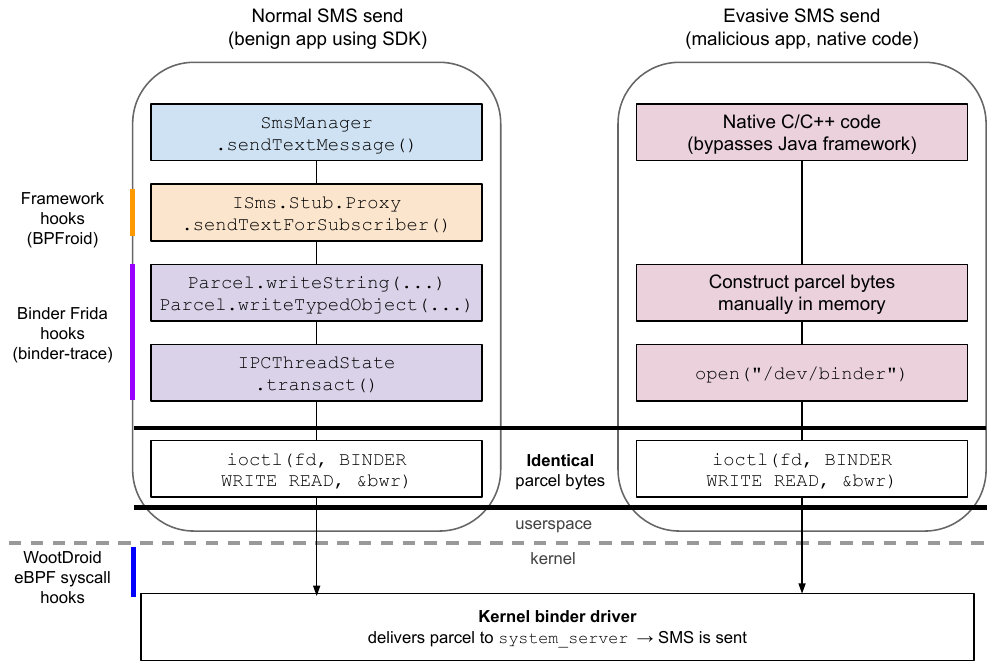}
    \caption{Two paths for sending an SMS via Binder and three
    instrumentation vantage points. Left: a benign app uses the
    Android SDK, passing through the framework and \texttt{libbinder}.
    Right: an evasive app constructs the parcel in native
    code~\cite{troopers2023beyondjava} and calls \texttt{ioctl}
    directly, bypassing the framework. Both paths produce identical
    bytes at the \texttt{ioctl} boundary.
    BPFroid's~\cite{BPFroid} eBPF uprobes on framework \gls{API} methods
    and binder-trace's~\cite{binder-trace} Frida hooks in
    \texttt{libbinder} instrument the left path only and are absent
    on the right. \WDBind captures both paths at
    the kernel boundary.}
    \label{fig:motivation}
\end{figure}

\point{The semantic gap at the \syscall layer}
A \syscall-only tracer, such as a port of \texttt{auditd} or a
straightforward application of \texttt{ftrace}, observes an SMS
send as a sequence of generic operations---an
\texttt{ioctl(BINDER\_WRITE\_READ)} on a file descriptor pointing
to \texttt{/dev/binder}, preceded and followed by unrelated reads
and writes. None of these events carries the destination number,
the message body, or the name of the Android \gls{API} being invoked;
the payload is a binary \texttt{Parcel} buffer whose layout is
determined by the Android framework's
\texttt{ISms.sendTextForSubscriber} interface description, not by
any information visible in the \syscall arguments. Existing work
addresses this semantic gap in one of several ways, each with a
limitation \WD aims to avoid.
ClearScope~\cite{ClearScope} is implemented as a custom build of the Android operating
system, both a modified runtime and an instrumented copy of the
platform code, and is therefore not deployable on a stock
smartphone, which precludes the use case \WD targets.
CopperDroid~\cite{CopperDroid} relies on Virtual Machine Introspection and
decodes parcels off-device in a separate oracle, precluding
on-device online analysis.
SliceDroid~\cite{alexopoulos2025slicedroid} recovers only
coarse-grained information such as which device node was accessed,
without identifying the invoked method or its arguments.
Binder-trace~\cite{binder-trace} and BPFroid~\cite{BPFroid}
rely on userspace instrumentation of the target application
with Frida and eBPF uprobes respectively.

The latter two approaches, in particular, are fragile in the
presence of an application that is aware of being traced. To
illustrate the problem, consider
the example of \cref{fig:motivation}.
A hypothetical Android application is distributed
through a third-party store as a PDF reader. Its outer \gls{APK}
contains only benign rendering code and defeats static scanning;
the behavior of interest is delivered post-install as a native
shared library, a dropper architecture that recent
threat-intelligence reporting identifies as the dominant delivery
pattern for SMS-abusing Android malware observed in
2025~\cite{groupib2025wonderland}. The payload performs
anti-instrumentation checks on startup and terminates if Frida,
Xposed, or common userland hooking frameworks are detected, a
technique documented in the WeddingCake packer observed in
thousands of Chamois-family samples~\cite{vb2018weddingcake} and
common in contemporary Android
crypters~\cite{ruggia2025darkside}. When such an application
eventually sends an SMS, it does so through a code path chosen
specifically to avoid user space instrumentation: rather than
calling \texttt{android.telephony.SmsManager}, which would route
through \texttt{framework.jar}, JNI, and \texttt{libbinder.so}, the
native library opens \texttt{/dev/binder} directly, marshals the
parcel for \texttt{ISms.sendTextForSubscriber} in its own address
space, and issues \texttt{ioctl(BINDER\_WRITE\_READ)} with a
hand-constructed transaction. This technique has been publicly
demonstrated with proof-of-concept code that rewrites an Android
application entirely in native code and invokes Android framework
services through direct Binder transactions rather than through
\texttt{libbinder}~\cite{troopers2023beyondjava}; it requires no
privileges beyond those of an ordinary app. A Frida-based tracer
such as binder-trace observes nothing in this scenario, for two
independent reasons: the process does not link \texttt{libbinder}
and therefore exposes no hook points in that library, and the
anti-instrumentation check terminates the payload if Frida is
attached at all.
Similarly, a uprobe-based tracer such as BPFroid also observes nothing,
as the application does not utilize the \glspl{API} of the Android framework. 

We emphasize that this scenario is hypothetical; we have not
evaluated \WD against a specific evasive sample. It is included
to make explicit a limitation of user space-instrumenting tracers
that applies in principle to any application aware of its
environment. The broader design point is that on-device,
semantically complete reconstruction of Binder transactions should
not depend on instrumentation that the traced application can
reach. \WD observes the parcel as the
kernel copies it on transaction delivery, a vantage point below
any user space hook.

\point{Event loss in \syscall tracing}
The second shortcoming is independent of the semantic gap and
affects the \syscall channel directly. The operations that
commonly surround a Binder transaction in a realistic audit
scenario, i.e.\ the \texttt{openat} that materializes a payload, the
\texttt{connect} or \texttt{sendto} that reaches a remote
endpoint, are \syscalls whose arguments \emph{are} meaningful at
the kernel level, provided the tracer delivers them reliably. In
practice, tracing pipelines based on \texttt{ftrace} 
rely on per-CPU ring buffers drained by a user space
reader, and when events are produced faster than the reader can
drain them, older entries are silently overwritten.
For auditing this is a
correctness concern rather than a performance concern: a single
missed \texttt{connect} breaks the causal chain from a suspicious
payload fetch to its destination. \textcite{eAudit} demonstrate
this limitation on Linux hosts and propose an \ebpf-based tracer
that substantially reduces event loss under load. Whether a
similar approach is viable on Android, given the platform's
restrictions on \ebpf and its different workload characteristics,
is an open question that \WD answers in the affirmative.

\point{What a complete audit record requires}
An audit system aiming to capture the SMS scenario faithfully
must therefore address both shortcomings simultaneously. It must
observe the Binder transaction at a vantage point where the
parcel is available and decode it semantically, so that the record
names the \gls{API} being invoked and its arguments. It must also
observe the surrounding \syscall activity without dropping events
under load, so that the record is complete.
Existing
approaches address at most one of these requirements: user space
Binder tracers recover semantics but leave the \syscall channel to
separate, unreliable tooling; \syscall-based provenance systems
recover the \syscall channel but stop at the Binder
\texttt{ioctl}. \WD combines both, at the kernel level and
without user space instrumentation of target \glspl{app}, which is the central claim of
this paper and the basis for the design we present in
\cref{sec:approach}.

%% file: content/04-approach.tex
\begin{figure}
    \centering
    \includegraphics[width=0.9\textwidth]{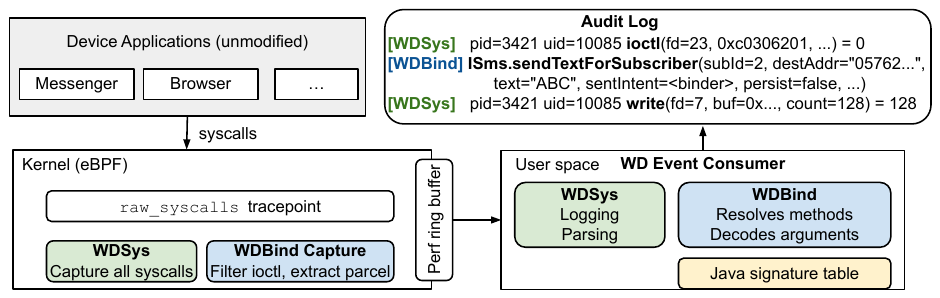}
    \caption{Architecture of \WD. Applications run unmodified.
    Two eBPF programs in the kernel
    (\WDSys and \WDBind) attach to the \texttt{raw\_syscalls}
    tracepoint and emit events through a perf ring buffer to a
    separate userspace consumer process. The consumer formats
    syscall log entries (\WDSys) and resolves method names and
    typed arguments for captured Binder transactions (\WDBind),
    using a signature table extracted ahead of time via Java
    reflection. The resulting audit log interleaves \syscall-level
    events with semantically decoded Binder calls.} 
    \label{fig:wootdroid}
\end{figure}

\section{Approach: \WD}
\label{sec:approach}
\WD comprises a user-space component for trace configuration and collecting the necessary method
mappings, and two eBPF components that jointly implement the main tracing logic, as shown in \cref{fig:wootdroid}.
In the following, we present more details on our approach.
Specifically, we present \textit{\WDSys}, the \syscall logging component (\cref{sec:approach-syscall}), and \textit{\WDBind}, which parses Binder transactions and reconstructs \gls{IPC} calls including their parameters (\cref{sec:approach-binder}).
In the end, we cover details of our implementation (\cref{sec:wootdroid-implementation}).

\subsection{\Glsentrylong{syscall} Tracing (\glsentryshort{WDSys})}
\label{sec:approach-syscall}
\gls{WDSys} adapts the state-of-the-art eAudit approach (see \cref{sec:background-eAudit}) for use in Android devices.
Adapting eAudit to Android is not trivial, as it requires overcoming several challenges, described below.

\begin{table}
    \centering
    \caption{Comparison of \syscalls traced (a) by both eAudit \cite[Table 1]{eAudit} on x86\_64 architectures and \WDSys on arm64 architectures, {\color{orange}(b) only with x86\_64/eAudit but not arm64/\WDSys}, and {\color{blue}(c) only with arm64/\WDSys but not x86\_64/eAudit}.}
    \footnotesize{
    \begin{tabular}{|p{\linewidth}|}
        \hline
        accept accept4 bind chdir {\color{orange}chmod} clone {\color{orange}clone3} close {\color{orange}creat} dup
        {\color{orange}dup2} dup3 execve execveat exit {\color{orange}exit\_group} fchdir fchmod fchmodat
        fcntl finit\_module {\color{orange}fork} ftruncate getpeername init\_module kill {\color{orange}link} linkat
        {\color{orange}mkdir} mkdirat {\color{orange}mknod} mknodat mmap mprotect {\color{orange}open} openat {\color{orange}pipe}
        pipe2 {\color{orange} pread} pread64 preadv {\color{blue}preadv2} ptrace {\color{orange}pwrite} pwrite64 pwritev {\color{blue}pwritev2} read
        readv recvfrom recvmmsg recvmsg {\color{orange}rename renameat} renameat2 {\color{orange}rmdir}
        sendmmsg sendmsg sendto setfsgid setfsuid setgid setregid setresgid
        setresuid setreuid  setuid socket socketpair splice {\color{orange}symlink} symlinkat tee
        tgkill tkill truncate {\color{orange}unlink} unlinkat {\color{orange}vfork} vmsplice write writev\\
        \hline
    \end{tabular}}
    \label{tab:syscall_traced}
\end{table}

\point{Tracepoint Availability}
\label{sec:approach-syscall-challenges-tracepoints}
Individual \syscall tracepoints are unavailable on recent Android \gls{OS} smartphones.\footnote{We specifically tested Google Pixel 7 and 9.}
Therefore, \WD uses the raw \texttt{sys\_enter} and \texttt{sys\_exit} tracepoints of the group \texttt{raw\_syscalls}.
Modifications to eCap were made such that specific operations per \syscall are executed based on switching between the \syscall id of \texttt{sys\_enter} for the enter of \syscalls and \texttt{sys\_exit} upon the return of \syscalls.
Additionally, structs for each of the \syscalls were designed in order to be able to utilize the functionalities of the rest of eCap without heavy modification.

\smallskip\point{Complexity}
Since using raw \syscall tracepoints increases the complexity of the \ebpf program, we additionally
had to split tracing logic across multiple programs, chained via tail calls, in order to circumvent stack limitations of the \ebpf verifier.
To achieve this, we created a program array with entries to link to individual programs.
We cross-compiled eAudit for Android using Extended Android Tools, which required some additional modifications. For example some imports caused errors on Android---instead we hard coded constants.

\smallskip\point{Architectural Differences}
Architectural difference between desktop x86\_64 and mobile arm64, meant that we needed to trace different \syscalls than eAudit did for desktop. \cref{tab:syscall_traced} shows the list of traced \syscalls and the architectural differences. While eAudit for desktop x86\_64 traces 81\footnote{eAudit actually reports only 80 \syscalls being traced in Table 1 \cite{eAudit}. This is likely because eAudit also traces the fcntl \syscall, but only dup operations are considered.} \syscalls, \WDSys for mobile arm64 only needs to trace 64 \syscalls, since 19 \syscalls being traced on x86\_64 are not available on arm64. Instead two additional newer \syscalls are added and traced, resulting in a total of 64 \syscalls being traced by \WDSys.

\subsection{Binder Tracing (\glsentryshort{WDBind})}
\label{sec:approach-binder}
In order to overcome the semantic gap problem commonly encountered when solely relying on \syscall logs, our approach additionally parses the ioctl \syscall buffer contents to record Binder transaction details.
Our approach adapts to \ebpf and integrates into the eAudit design, established Binder tracing and parsing logic~\cite{CopperDroid,btrace,binder-trace}, facilitating practical on-device \gls{API} invocation reconstruction without user space instrumentation.

\begin{figure}
    \centering
    \includegraphics[width=\linewidth]{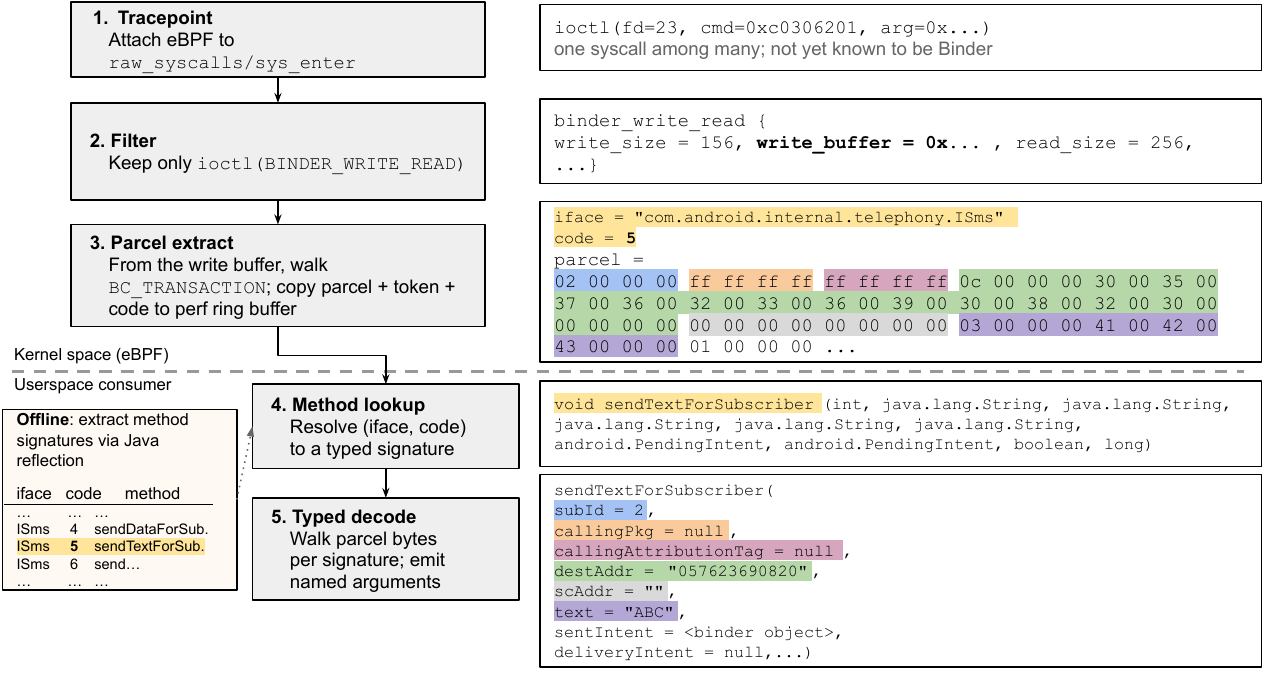}
    \caption{Overview of \WDBind with Binder transaction parsing of SMS send example.}
    \label{fig:wootdroid-binder}
\end{figure}

An overview of how \WDBind works is shown in \cref{fig:wootdroid-binder}. 
\textbf{\Circled{0}} Beforehand, we obtain a method signature table extracted using Java reflection directly from the Android framework.
\textbf{\Circled{1}} \WDBind attaches an \ebpf tracepoint to capture \syscalls.
\textbf{\Circled{2}} We parse and filter \syscalls to only consider \texttt{ioctl} \syscalls with the \texttt{BINDER\_WRITE\_READ} command code.
\textbf{\Circled{3}} Through multiple structures we extract the contained Binder parcels, interface token, and method code for Binder transactions as indicated by the \texttt{BC\_TRANSACTION} Binder command.
Data on these transactions is passed from the \ebpf program in kernel space to the \WD user space consumer, which consumes the events for further processing.
\textbf{\Circled{4}} Subsequently, The \WD consumer resolves the interface and code of transaction events to a typed signature containing information about the arguments of the specific \gls{API} call, by consulting the previously extracted signature table.
\textbf{\Circled{5}} Finally, we are able to parse the parcel bytes using the previous information on argument types.
In the following, we provide more details on each step.

\point{\Circled{0} Offline Method Mapping}
\label{sec:approach-binder-mapping}
In order to understand buffer content, we need a way to gather information about the specific method called and its parameters. Similar to btrace \cite{btrace}, we extract a Java signature table from codes to method names offline from the device for all possible classes and interfaces. Additionally, we derive details about the parameters per methods. In particular, we use a companion app that is run offline before any tracing to dump this table. At the same time, we obtain a list of such classes by enumerating running system services on the device in question.

Specifically, using Java reflection, we obtain the specific \texttt{\$Stub} class for a interface in question.
We iterate over the declared fields of the class, whereby the index corresponds to the code argument contained in Binder transactions. We find the corresponding method names of each index by using \texttt{Binder\-Transaction\-Name\-Resolver}'s \texttt{getMethodName()} and match them with the actual method of the class. Once we found it, we can then enumerate the arguments and their types of the method using \texttt{getParameterTypes()} and build the Java signature table.

\point{\Circled{1} Tracepoint \& \Circled{2} Filter}
In order to capture Binder transactions, our \ebpf program in the kernel first captures \texttt{ioctl} \syscalls by tracing the \ebpf \texttt{raw\_syscalls} tracepoint and filtering out all other occurring \syscalls. The workflow is based upon the mechanisms employed by the Android system to parse Binder transactions as performed by \texttt{binder\_ioctl\_write\_read()} \cite{BindercAndroidCode}, but following this in \ebpf.
This means, we are mainly interested in tracing the BINDER\_WRITE\_READ command code (see \cref{sec:background-binder}).

\point{\Circled{3} Parcel Extract}
For each such encountered \texttt{ioctl}, our \ebpf program parses the write of the \texttt{binder\_write\_read} structure passed as argument to the \syscall.
Following that, we iterate over commands contained in the buffers. Here we focus on \texttt{BC\_TRANSACTION} commands. Following each command in the buffer is their associated data content. We are especially interested in \texttt{BC\_TRANSACTION} commands, for which associated data is provided as a \texttt{binder\_transaction\_data} structure. This structure in turn also contains a \texttt{buffer} containing different Binder objects,
In the most common case of \texttt{BC\_TRANSACTION}, the buffer contains a parcel expressing a method call. The beginning of the buffer contains 12 bytes of header information that we skip while parsing, followed by the length of the interface token (4 bytes), a string of the interface token (16 bytes) and parameters of the method call.
In other cases, this might include arbitrary flat binder objects that we further parse.
After successful capturing of Binder transactions in \ebpf kernel space, these events are passed to the \WD user space consumer.

\point{\Circled{4} Method Lookup}
\label{sec:approach-binder-parsing}
For each Binder transaction event, the WD consumer in user space receives the interface token, code and buffer content. For a full list of our event record see \cref{apx:approach-binder-record}.
Using the token and code, we can fetch the corresponding method signature from the mapping of the initial offline step, thus obtaining method name, and details about the parameters including types.

\point{\Circled{5} Typed Decode}
Subsequently, this information enables us to parse the remainder of the Parcel's data and obtain complete information on a specific Binder transaction.
We iteratively parse each parameter depending on its data type and advancing our pointer on the buffer by each type's size:
Primitives, i.e., int, boolean, double, or long are simply read according to their size (4 or 8 bytes).
\texttt{java.lang.String} are handled according to \texttt{Parcel.writeString()}, meaning \WDBind reads the prefixed length first, followed by a UTF-16LE encoded string, a null terminator, and padding.

More complex, flat Binder objects are correctly recognized and their data contained in the Parcel parsed. 
We employ a simple heuristic to differentiate between objects like \texttt{IBinder} written via \texttt{writeStrongBinder()} vs. objects like \texttt{PendingIntent} written via \texttt{writeTypedObject()}.
The latter necessitates an additional check beforehand for a nullness marker, to indicate whether an object will follow or simply a \texttt{null} value was passed.
An example for such a transaction being captured and parsed by \WD is provided in \cref{fig:wootdroid-binder}.

\point{Benefits}
In general, we make sure to adjust for potential byte alignment after each parameter.
Usually parsing of parcel data would be guided by the \texttt{offset} buffer contained in the transaction, however, due to complexity limitations with \ebpf, as later described, we opted for our proposed approach.

While we do recognize and parse primitive and flat Binder objects, we do not follow file descriptor or pointer references and do not resolve higher levels of indirection currently.
However, in contrast to related works, we do so without relying on an external Oracle like CopperDroid \cite{CopperDroid} does.
Additionally, we also do not rely on previously generated structs generated from the Android source code like \cite{binder-trace}, but instead obtain the ground truth directly from the device in question.

\subsection{Implementation Details}
\label{sec:wootdroid-implementation}

\point{\Glsentrylong{eBPF} on Android}
\label{sec:AndroideBPF}
Since Android is based on Linux, it supports and uses \ebpf \cite{ExtendKernelEBPF}, for example, for the networking daemon (netd), power estimate calculation based on CPU statistics and GPU memory profiling. However, the function range out of the box is limited since access to \ebpf is heavily restricted using SELinux and Android's BPF loader, the types of \ebpf programs are limited, user space tooling is missing, and deployment is difficult while passing device integrity checks by Google. Especially older smartphones, Android versions, and corresponding kernel versions lack basic features necessary to easily and efficiently run \ebpf programs.

Using efficient perf ring buffers in \WD necessitates a rather new kernel version as this feature is only supported since Linux 5.8 kernel. This means that at minimum Google Pixel 6 \cite{BuildPixelKernels} is required, since previous Pixels only support up to kernel 4.19 officially.
To run \ebpf programs, we use the \gls{BCC} \cite{IovisorBcc2026} Python library.
To set up an environment on Android, we utilize the "Extended Android Tools"\cite{FacebookexperimentalExtendedAndroidTools2026} which provide capabilities to cross-compile common Linux tools for Android, including \gls{BCC} and Python. Alternative approaches \cite{joelJoelagnelAdeb2026,weishuTiannEadb2026} also exist and work by deploying a chroot-like environment on top of the normal \gls{adb} shell.

\point{\ebpf Features} 
In order to fulfill \ebpf complexity requirements for \WDBind, we make use of tail calls, hash maps, and the perf ring buffer:
We again split functionality across multiple chained tail calls to reset stack limitations.
We save encountered results and addresses into hash maps for subsequent processing across tail calls.
We use a perf ring buffer \cite{BPFRingBuffer} to output events back to our user space event consumer.
Additionally, buffer reads are split up in chunks for larger reads, meaning multiple \ebpf events might be passed to the user space, where they are stitched together again.

\point{Memory Access}
Starting with Android 13, supported smartphones like Google's Pixel 8 \cite{ArmMemoryTagging} and newer will use \gls{MTE} \cite{ArmMemoryTagginga} to protect against memory safety bugs.
As such, user space memory read operations with \ebpf on recent Android versions require a sparsely documented workaround in order to be able to access user memory correctly. Because \WDBind needs to read user space memory for extracting parcel data, for such addresses a bitwise AND with \texttt{0xffffffffff} is calculated. 

\point{Deployment and Rooting}
Tracing with \ebpf commonly requires privileged access, which is why the smartphones used for evaluation are using stock Android with the only exception of being rooted, meaning the boot partition was modified using Magisk\cite{wuTopjohnwuMagisk2026} in order to provide the common \texttt{su} binary to access superuser privileges.
Alternative solutions exist like pre-loading the \ebpf code of our approach into the system \ebpf startup programs \cite{ExtendKernelEBPF}. But these would require either a modification of the Android system or a custom kernel with particular configuration options.
Adding our approach to the Android system and automatically loading it upon boot is possible and would be the ideal and preferred solution. However, this is only realistically and transparently feasible, and has been done already, in the role of a phone manufacturer that can sign the system images correctly%
~\cite{eBPFXiaomi}.

%% file: content/05-evaluation_results.tex
\section{Evaluation}
\label{sec:eval}
We stipulate the following research question, which will be answered through the evaluation of our approach:

\begin{itemize}
    \item[RQ1] What \textbf{performance impact} does our tracing with \WDSys have on the mobile system in comparison to other approaches? (\cref{sec:eval-performance})
    \item[RQ2] How \textbf{complete} is the information we are able to trace with \WDSys in comparison to other approaches? (\cref{sec:eval-completeness})
    \item[RQ3] Can \WDBind \textbf{reconstruct} security-relevant application behaviors? (\cref{sec:eval-binder-casestudy})
\end{itemize}

\renewcommand\theadfont{\bfseries}
\renewcommand{\cellalign}{lc}
\renewcommand{\theadalign}{lc}
\begin{table}
    \centering
    \ra{1.2}
    \caption{Overview of evaluation experiments.}
    \begin{tabular}{l|l|l}
        \toprule
        \thead{RQ} & \thead{Approach} & \thead{App under Test}\\ \midrule
        RQ1: Performance & A1 \WDSys vs. A2 ftrace & Geekbench\\
        RQ2: Completeness & A1 \WDSys \& A2 ftrace & Top 100 GPlay Apps\\
        RQ3: Case Study & \WDBind & Whole-system\\
        \bottomrule
    \end{tabular}
    \label{tab:eval-overview}
\end{table}

\subsection{Overview and Setup}
\label{sec:eval_setup}

To bridge the semantic gap problem, \WD enables two-fold trace collection, capturing \syscall logs and Binder transactions logs, by running \WDSys and \WDBind. 
We compare \WDSys results with the equivalent \syscall approach from related works, namely the general-purpose Linux-specific approach for \syscall tracing called ftrace, to answer RQ1 and RQ2, while providing a case study on \WDBind's reconstruction capabilities. 
We exclude BPFroid \cite{BPFroid} and AuditDOnAndroid \cite{AuditDAndroid} since we were not able to run them on recent Android versions. 
In summary, \cref{tab:eval-overview} shows an overview of experiments we run for evaluating our \WD approach.

We run experiments on two Google Pixel~9 smartphones, running Android~16, rooted using Magisk.\footnote{For the reason why rooting is necessary in our setting and what is needed to avoid it, please refer to the end of \cref{sec:AndroideBPF}.} We automate the execution of applications depending on their use case, as described in the following sections for RQ1 and RQ2 each.
Further details on the deployment of \ebpf were already covered in \cref{sec:AndroideBPF}.

\point{Workflow}
To run our evaluation, we created a framework that enables automation and reproducibility:
\begin{enumerate}
    \item \textbf{Device Setup}: flashing different Android stock images for Pixel smartphone, rooting these with Magisk.
    \item \textbf{Approach Setup}: creation of executable archives for \gls{BCC} and eAudit for Android using cross-compilation and necessary modifications.
    \item \textbf{Tracers}: Automation for setup, configuration, execution, and log fetching.
    \item \textbf{Applications}: Support the gathering and execution of different kinds of application and instrumentation strategies.
    \item \textbf{Experiments}: Scripting and configuration for the execution of different experiments, entailing a set of tracers, a set of applications and automated app interaction.
    \item \textbf{Scenarios}: (a) Executing common performance benchmark \glspl{app}, capturing metrics and evaluation thereof. (b) Executing \glspl{app} from the Google Play store with automated \gls{app} interaction, and behavior analysis by investigating differences in completeness of tracer logs.
\end{enumerate}

\subsection{RQ1: Performance}
\label{sec:eval-performance}
To validate our requirements and answer our research questions, we run multiple experiments to measure performance impact of the logging approaches. We define performance impact by measuring system performance as judged by benchmarks and comparing this baseline to executions with specific logging approaches enabled.

\point{Performance Benchmarks}
\label{sec:eval-rq1-benchmarks}
Similar to the work of \textcite{DBLP:conf/uss/0003WLMG24}, we run the Geekbench suite \cite{Geekbench6CrossPlatform} to measure CPU performance and repeat experiments ten times to produce reliable results. Benchmarks execution is automated using AndroidViewClient \cite{milanoDtmilanoAndroidViewClient2026}.
Geekbench includes scores for single-core and multi-core performance, and GPU performance scores. Higher scores mean better performance.

\label{sec:eval-rq1a-performance-syscall}
\begin{table}
    \centering
    \ra{1.2}
    \caption{Tracing overhead of \WDSys and ftrace (Geekbench6 scores).}
    \begin{tabular}{ll|ll|l}
        \toprule
        \thead{Bench.} & \thead{Baseline} & \thead{Tracer} &  \thead{w/Tracing} & \thead{Overhead} \\ \midrule
        \multirowcell{2}{Single\\Core} & 1487.2 & A1 \WDSys & 1433 & 3.6\% \\
         & 1487.2 & A2 ftrace  & 1399.1 & 5.9\% \\ \hline 
        \multirowcell{2}{Multi\\Core} & 3651.4 & A1 \WDSys & 3618 & 0.9\% \\
         & 3651.4 & A2 ftrace & 3545.6 &  2.9\%\\
        \bottomrule
    \end{tabular}
    \label{tab:eval-rq1a}
\end{table}

\point{Performance Results}
\cref{tab:eval-rq1a} summarizes the benchmark results of our baseline, \WDSys, and the compared approach ftrace. 
Tracing with \WDSys incurs only minimal runtime overhead of 3.6\% and 0.9\% on Geekbench6.
In comparison, ftrace has a slightly higher overhead of 5.9\% and 2.9\%.
Therefore, \WDSys is 2-2.3\% more efficient in terms of performance than ftrace.
While we also measured the impact of tracing on the GPU benchmarks, we do not report them in further details, since the overhead is negligible, making up at most 0.5\% difference in scores. This is to be expected since tracing is typically performed using the processor and not utilizing the GPU.

\subsection{RQ2: Completeness}
\label{sec:eval-completeness}
We showcase the utility of understanding application behavior by investigating the execution of multiple Android applications. Here, we compare the completeness of produced logs of different approaches to ours. For this, we run logging approaches on the same device in parallel to capture the same behavior. We create a dataset of the most used 100 applications from the Google Play store and run each \gls{app} while stimulating interaction using Android's UI/Application Exerciser Monkey \cite{UIApplicationExerciserb}, injecting 1000 inputs at an interval of 500ms.
While we acknowledge that its use does not provide complete coverage of \glspl{app} behavior, similar to other works \cite{NisiReconstruction}, we select Monkey for this purpose, since high coverage of \glspl{app} behavior is not relevant to the goals of this study.

\point{App Dataset}
\label{sec:eval-rq2-dataset}
To create our dataset, we first query the Google Play store for the top 100 free \glspl{app} across all categories using google-play-scraper \cite{olanoFacundoolanoGoogleplayscraper2026}.
We then download the \glspl{APK} of these \glspl{app} from the Google Play store using apkeep \cite{EFForgApkeep2026}.
Here, also split \glspl{APK} \cite{BuildMultipleAPKs} are considered and downloaded.
This is necessary because \glspl{APK} of applications nowadays are usually split in order to reduce their size and create specific configurations for different device configurations, i.e., screen density, architecture, or \gls{ABI}.
For an overview of the applications contained in our dataset, refer to \cref{apx:dataset-completeness-full}.

\point{Execution}
To evaluate the completeness of our \syscall (audit) tracing, we activate tracing of the two \syscall auditing approaches A1 \WDSys and A2 ftrace in parallel. Since they trace the whole-system, we make sure to ignore relevant pids related to the other tracer execution. We do so by first starting A2 to obtain its pids, then starting A1 and passing A2's pids to ignore, and finally update A2 to ignore A1's pids as well.
To cover a wide range of executions, we execute each app of our GPlay dataset with Monkey.
Each application run produces two logs of A1 and A2 respectively.
We compare both logs for each application run and investigate the completeness of contained \syscall events in both.

\point{Log Normalization}
Before we can compare events between the two tracers, we need to normalize both logs into a common format.
Of particular note is the timestamp of \syscall events, since \WDSys uses absolute timestamps and ftrace uses timestamps relative to the boot time.
Because ftrace logs \syscall enter and exit events separately and \WDSys logs them jointly, we combine matching enter-exit events.
Further normalization steps included additional pre-processing like filtering tracers' pids and harmonizing data representations.

\point{Event Matching}
Since unique identifiers for \syscall events are not available from the tracers' logs, we apply an algorithm to match \syscall events between the two tracers.
Matching of events is performed by matching the tgid or pid, \syscall id and \syscall arguments, as follows:
For each \syscall event in A2's log (ftrace), we look for a matching event with the same \syscall and pid in A1's log (\WDSys) that has not yet been matched.
Based on the specific \syscall, we then compare and look for the first event where the remaining arguments of the particular \syscall match.
We only consider the time frame where both tracers are actively logging since they are started sequentially, as described previously.
For calculating the initial timestamp offset between the two tracers, we look for the first matching \texttt{mmap} \syscall following an \texttt{execve} syscall.
Additionally, we ignore \texttt{clone} \syscalls since there is no way to distinctively match them. 

\point{Completeness Metric}
Once matching of events is completed, we gather statistics about the completeness of both tracers, by comparing the matching and non-matching events of the tracers with each other.
Of particular interest is the ratio of \emph{unique} events of one tracer in comparison to
all observed events of both tracers.
We define this \emph{unique event rate (UER)} in percentage as our completeness metric
to evaluate the performance of \WDSys since it expresses how much more one tracer is able to capture compared to the other.
Consider an example of a total of 100 events captured by both tracers (union).
If tracer A has a UER of 0.5 and tracer B has a UER of 0.1, this means that
A captured 50 unique events that B did not capture and B captured
10 unique events that A did not capture. In total, A captured 90 events and B captured 50 events.

\label{sec:eval-rq2a-completereness-syscall}
\begin{figure}
    \centering
    \includegraphics[width=0.75\linewidth]{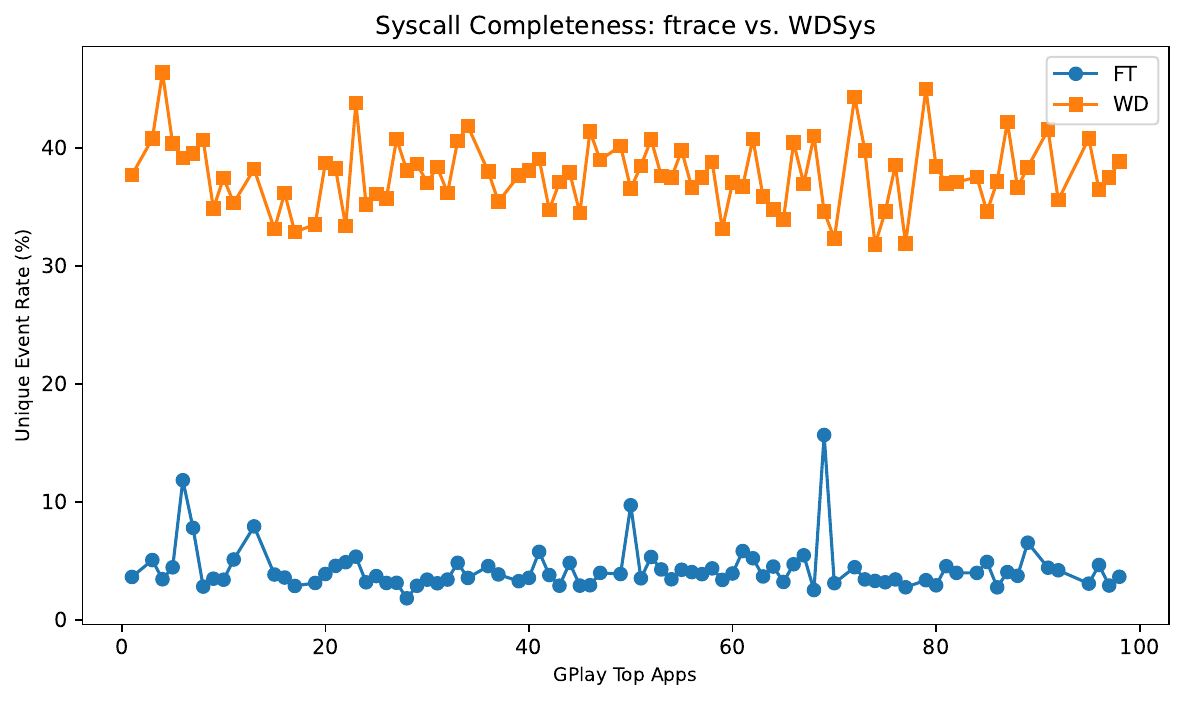}
    \caption{Unique event rates (UER) of \WDSys and ftrace across 100 application runs.}
    \label{fig:eval-rq2a}
\end{figure}

\point{Completeness Results}
\cref{fig:eval-rq2a} shows the completeness metric of both tracers for each application run.
Complete results are provided in \cref{apx:dataset-completeness-full}.
UER for \WDSys ranges from 31.81\% to 46.37\%, while ftrace only achieves a UER between 1.94\% and 15.68\%.
The mean UER for \WDSys is 37.75\% with a standard deviation of 2.92\%. In contrast, the mean UER of ftrace is 4.27\% with a standard deviation of 1.94\%. We can clearly see that \WDSys outperforms ftrace, since \WDSys systematically observes significantly more unique events.
Specifically, \WDSys on average captures 33.48\% (37.75-4.27) more events than ftrace.

\subsection{RQ3: Binder Case Study}
\label{sec:eval-binder-casestudy}
In the following, we present a case study on the reconstruction of Binder transactions pertaining to Android framework method calls using our \WDBind approach.
\begin{table}
    \centering
    \ra{1.2}
    \caption{Binder Case Study.}
    \begingroup\footnotesize
    \begin{tabular}{lp{3cm}p{6cm}}
        \toprule
        \thead{Class} & \thead{Method} &  \thead{Parameters} \\ \midrule
        ISms & sendTextFor\newline Subscriber & \texttt{[2, null, null, "057623690820", "", "ABC", ..., 0x5c27edfc88bdfc8b]} \\ 
        ITelephony & getDeviceIdsForUid & \texttt{[10180]}  \\
        ITelephony & isEmergencyNumber & \texttt{["123456789", 1]} \\ \hline 
        IPackageM. & getPackageInfo & \texttt{["com.google.android.apps.turbo", 512, 0]} \\ 
        IPackageM. & getApplicationInfo & \texttt{ ["com.google.android.dialer", 268436480, 0]} \\
        IPackageM. & getPackageUid & \texttt{["com.google.android.gms", 0, 0]} \\
        IPackageM. & getInstallerPackage Name & \texttt{"com.google.android.apps.messaging"} \\
        IActivityM. & checkPermissionFor Device & \texttt{["android.permission.READ\_CALL\_LOG", 23376, 10180, 0]} \\ \hline
        INotificationM. & cancelNotification\newline WithTag & \texttt{["com.google.android.youtube", "com.google.android.youtube", null, 1005, 0]} \\
        IAccountM. & getAccountsAsUser & \texttt{["com.google", 0, "com.google.android.gms"]} \\
        \bottomrule
    \end{tabular}
    \endgroup
    \label{tab:eval-binder-casestudy}
\end{table}

Based on SuSi \cite{DBLP:conf/ndss/RasthoferAB14}, we selected ten framework methods of security-relevant behaviors for our case study, covering different categories. These examples of \gls{IPC} calls and our (abbreviated) produced audit record are shown in \cref{tab:eval-binder-casestudy}. The corresponding traces were collected by manually interacting with different \glspl{app} on the phone, including the Messages, Phone, and YouTube \gls{app}.
These methods are important because they reveal sensitive information related to phone calls (\texttt{ITelephony} class) and information about installed \glspl{app} including data on their usage patterns like permission use (\texttt{IPackageManager.*}). Observability into which accounts an app enumerates (\texttt{IAccountManager.getAccountsAsUser}) or which notification apps clear (\texttt{INotificationManager}) are relevant events our approach captures.  Especially considering our motivational example of a potential evasive app (\cref{sec:motivational_example}), \WD enables checking whether such calls are circumvented by using native code and not visible to common approaches relying on \gls{API} call tracing. 
Looking at the \texttt{checkPermissionForDevice} trace example, we can observe that \texttt{READ\_CALL\_LOG} permission was requested by the Phone \gls{app} (as indicated by pid 23376 and uid 10180).
From the \texttt{isEmergencyNumber} call we can correctly identify that the phone number 123456789 was called.

\point{SMS Send Example}
Coming back to our motivating example (\cref{sec:motivational_example}), when an application performs an SMS send via \texttt{ISms.sendTextForSubscriber}, we observe the audit record of \cref{fig:wdbind-sms-record}. For the full details on the buffer of this Binder transaction, refer to \cref{apx:binder-casestudy-additional}.
We see that \WDBind correctly reconstructs primitive data types like int, boolean, and long.
Additionally, it handles flat Binder objects like \texttt{PendingIntent} when reading the parcel data. \texttt{PendingIntent} is passed as \texttt{BINDER\_TYPE\_HANDLE} and written via \texttt{writeTypedObject()}. Also, \WDBind correctly recognizes and handles the resulting nullness marker present in this case, while recognizing other interfaces like \texttt{IBinder} that are written directly via \texttt{writeStrongBinder()} without any markers.

\begin{figure}[h]
\centering
\begin{minipage}{\columnwidth}
\begingroup\footnotesize
\begin{verbatim}
pid=10119  uid=10188  flags=18  data_size=200 
iface=com.android.internal.telephony.ISms  code=5
sendTextForSubscriber(int subId=2, String callingPkg=null,
  String callingAttributionTag=null, String destAddr="057623690820",
  String scAddr="", String text="ABC",
  PendingIntent sentIntent=<HANDLE, flags=0x13, ..., stability=12>,
  PendingIntent deliveryIntent=null, boolean persistMessage...=1,
  long messageId=0x5c27edfc88bdfc8b)
\end{verbatim}
\endgroup
\end{minipage}
\caption{Audit record produced by WDBind for an SMS send. The record includes the calling process identity, the resolved method name and typed arguments}
\label{fig:wdbind-sms-record}
\end{figure}
\vspace{-0.5cm}

%% file: content/07-conclusion.tex
\section{Discussion and Conclusion}
\subsection{Discussion}
\label{sec:discussion}
\point{Kernel vantage point as a design principle}
\WD demonstrates that capturing Binder transactions at the \texttt{ioctl} boundary provides semantic reconstruction comparable to user-space hooking, without the evasion surface those hooks introduce. The parcel bytes must pass through the kernel regardless of how they were constructed, making the approach inherently robust to application-level anti-instrumentation, a property that holds even as evasion techniques in Android malware grow more sophisticated.

\point{Practical viability of eBPF on Android}
Our experience porting eAudit to Android reveals that while eBPF is nominally supported, significant Android-specific effort is needed to work around missing per-syscall tracepoints, limited BPF loader program types, MTE address masking, and the absence of standard user-space tooling such as BCC. While general eBPF challenges like verifier stack limits also apply, the combination of these Android-specific restrictions makes deploying custom eBPF programs considerably harder than on desktop Linux. These obstacles represent a practical barrier for security researchers. We hope that publishing our solution lowers this barrier for future Android eBPF tooling.

\point{Completeness metric and residual event loss}
The UER results show WDSys consistently captures substantially more unique events than ftrace across all tested apps, with low variance, validating eAudit's design for mobile environments. However, ftrace's non-zero UER (mean 4.27\%) indicates that WDSys also misses some events, which is noteworthy given that the underlying approach eAudit is specifically designed to minimize event loss.
Our completeness experiment setup can be used for further research on tracer evaluation and design.

\point{Signature table coverage and maintainability}
Because our method mapping is extracted via Java reflection directly from the device, it automatically adapts to any Android version without requiring AOSP source access. However, the signature table only covers AIDL-defined interfaces exposed through system services; vendor-specific or dynamically registered interfaces are left for future work.

\point{Complementarity of WDSys and WDBind}
Our two modules address orthogonal gaps in existing auditing: WDSys ensures the syscall record is complete, while WDBind ensures it is semantically meaningful. The interleaved audit log \WD produces is, to our knowledge, the first to combine both properties on unmodified Android.

\subsection{Limitations and Future Work}
\point{Requires root}
On current Android devices, root user access is required to deploy \ebpf programs, which is usually not available by default. Tools such as Magisk \cite{wuTopjohnwuMagisk2026} can be used to root a wide range of devices.
In the future, the ability to deploy the necessary eBPF tracing scripts could be offered by the Android system, as part of standard tracing utilities.

\point{Complex parameter types}
The current implementation of \WDBind is limited
to reconstructing method call parameters with primitive data types, Strings and some initial more complex Binder types in parcel data of Binder transactions.
It does not resolve these references further, does not handle Binder objects with file descriptors or pointers, and does not handle transaction replies.
Although primitive types and rudimentary support for more complex Binder types already cover a wide range of useful method calls, as shown in our case study,
we intend to include such functionality in future iterations of the system.

\point{Binder completeness and performance}
As stated initially, we only provided an end-to-end case study of reconstruction security-relevant behaviors, including our motivating example. 
While a fair comparison with user space-based frida tracing approaches is difficult and not straightforward, we already prepared the integration of other such Binder tracing approaches into our workflow.
A full systematic evaluation of the completeness and performance overhead of \WDBind reconstruction is, therefore, left as future work.

\subsection{Conclusion}\label{sec:conclusion}
We highlighted common limitations of existing system auditing approaches on Android.
In particular, popular \syscall tracers tend to lose events under load and Android-specific solutions rely on userspace hooks that are easy to detect and evade.
With \WD, we showed that security-relevant semantic information
can be collected from kernel hooks on unmodified devices.
Our contributions can assist investigations into evasive \glspl{app} behavior, but can also
serve as the basis for live intrusion detection and response systems.

%% file: content/xx-acknowledgements.tex
\begin{credits}
\subsubsection{\ackname}
    This research work was supported by the German Research Foundation (DFG) under grant number 251805230 (GRK 2050) and the National Research Center for Applied Cybersecurity ATHENE. ATHENE is funded jointly by the German Federal Ministry of Research, Technology and Space and the Hessian Ministry of Science and Research, Arts and Culture.

    
\end{credits}

%% file: content/xy-appendix.tex
\crefalias{section}{cref:appendix}
\crefalias{subsection}{cref:appendix}
\crefalias{subsubsection}{cref:appendix}
\crefalias{paragraph}{cref:appendix}
\crefalias{subparagraph}{cref:appendix}

\section{Appendix}
\label{apx:appendix}

\subsection{Background: Android Sandbox \& the Semantic Gap Problem}
Resources on Android are separated through multiple mechanisms, which Google refers to as Application Sandbox\footnote{\url{https://source.android.com/docs/security/app-sandbox}}. One such mechanism is the user ID-based discretionary access control. Hereby, each application is assigned a unique user ID such that apps cannot interact with each other and have limited access, based on plain and simple user separation of processes and file permissions. Additionally, another such mechanism called SELinux enforces mandatory access control to further isolate the system from apps and apps across user boundaries. SELinux runs in a special mode such that anything not explicitly allowed is denied.

This sandbox paradigm spawns another specialty of the Android system. Since \glspl{app} are strictly isolated from the system and from each other, they can only use certain functionalities and get certain information through other more privileged \glspl{app} or specifically provided system interfaces. In order to facilitate this exchange of information and communication between sandboxed applications Android 
\begin{enumerate*}[label=(\alph*)]
    \item runs different service managers with more privileged access to information like the location. \Glspl{app} need to communicate with each other and with such managers to utilize other \glspl{app} activities or access privileged system information via a specifically provided \gls{IPC} and \gls{RPC} mechanism. And
    \item additionally, Android provides high-level \glspl{API} for common activities and tasks, like networking or accessing user's information databases.
\end{enumerate*}

As a result, the semantics of \gls{app} activities remain at a higher level and do not pass through to the lower \syscall level. In other words, monitoring \syscalls only captures standard activities of standard processes, that actually perform those activities substitutionally for other \glspl{app}. The activities' root dependencies, as well as their intentions and contexts, remain hidden to the system level, which is referred to as the \emph{semantic gap problem}~\cite{NisiReconstruction}.

\subsection{Approach: \WDBind Event Record}
\label{apx:approach-binder-record}
In the following, we provide details on the attributes of the event record for Binder transactions events as traced by \WDBind.
\begin{table}[h]
    \centering
    \caption{Information obtained and logged per Binder transaction, whereby "logged" attributes are passed to user space and "obtained" attributes are only processed in kernel space.}
    \begin{tabular}{l|p{4.5cm}|p{4cm}}
        \toprule
        \thead{Category} & \thead{Attributes Logged} & \thead{Attributes Obtained} \\ \midrule
         Process &  ts, pid, uid & gid, comm \\
         ioctl & & fd, cmd, arg \\
         Binder write read &  & write\_size, read\_size, binder\_cmd \\
         Binder transaction & code, target flags, data\_size & handle, sender pid, offset\_size \\
         Parcel details & interface (token), (method) code, method name, reconstructed parameters &\\
         Raw data & (transaction) buffer  &\\ \bottomrule
    \end{tabular}
    \label{tab:event_record}
\end{table}

\subsection{Evaluation: Binder Case Study: Additional Materials}
\label{apx:binder-casestudy-additional}
In the following, we provide the full buffer of the studied SMS send transaction:

\begin{verbatim}
    raw_buffer:
     00 00 00 80 | ff ff ff ff  # header
     54 53 59 53                # header 
     23 00 00 00                # interface token length
     63 00 6f 00 | 6d 00 2e 00  # start interface token str
     61 00 6e 00 | 64 00 72 00 
     6f 00 69 00 | 64 00 2e 00
     69 00 6e 00 | 74 00 65 00
     72 00 6e 00 | 61 00 6c 00 
     2e 00 74 00 | 65 00 6c 00 
     65 00 70 00 | 68 00 6f 00 
     6e 00 79 00 | 2e 00 49 00 
     53 00 6d 00 | 73 00 00 00  # end interface token str
     
     02 00 00 00                # arg1: int subId = 2
     ff ff ff ff                # arg2: Str callingPackage = null
     ff ff ff ff                # arg3: callingAttributionTag=null
     0c 00 00 00                # arg4: (len 12)
     30 00 35 00 | 37 00 36 00  # arg4: destAddr = "057623690820"
     32 00 33 00 | 36 00 39 00  # arg4
     30 00 38 00 | 32 00 30 00  # arg4
     00 00 00 00                # arg4
     00 00 00 00 | 00 00 00 00  # arg5: scAddr = "" (len 0)
     03 00 00 00                # arg6: (len 3)
     41 00 42 00 | 43 00 00 00  # arg6: String text = "ABC"
                                # arg7: PendingIntent sentIntent
     01 00 00 00                # arg7: non-null marker 
     85 2a 68 73                # arg7: fbo type
     13 00 00 00                # arg7: fbo flags
     77 00 00 00 | 00 00 00 00  # arg7: fbo handle
     00 00 00 00 | 00 00 00 00  # arg7: fbo cookie
     0c 00 00 00                # arg7: fbo stability footer
     00 00 00 00                # arg8: P.I. deliveryIntent = null
     01 00 00 00                # arg9: bool persist... = true
     5c 27 ed fc | 88 bd fc 8b  #arg10: long messageId
 \end{verbatim}

\subsection{Evaluation: Full Completeness Results}
\label{apx:dataset-completeness-full}
In the following, we provide the completeness metric calculated for all applications from the top 100 Google Play Store app dataset used for our evaluation.
\begin{table}[h]
    \centering
    \caption{Full completeness results for dataset (top).}
    \begin{tabular}{l|l|l|l|l}
        \toprule
        \thead{\#} & \thead{Category} & \thead{App Package} & \thead{FT} & \thead{WD}\\
        \midrule
        1 & Auto \& Vehicles & de.kba.ikfz & 3.65 & 37.72 \\
        2 & Productivity & com.openai.chatgpt & \multicolumn{2}{l}{\textit{GPlay Err.}} \\
        3 & Auto \& Vehicles & de.fahrzeugschein.app & 5.08 & 40.78 \\
        4 & Shopping & com.einnovation.temu & 3.45 & 46.37 \\
        5 & Social & com.instagram.android & 4.46 & 40.37 \\
        6 & Video Players \& Editors & com.lemon.lvoverseas & 11.85 & 39.16 \\
        7 & Social & com.zhiliaoapp.musically & 7.81 & 39.55 \\
        8 & Libraries \& Demo & com.governikus.ausweisapp2 & 2.83 & 40.66 \\
        9 & Shopping & de.cellular.ottohybrid & 3.5 & 34.88 \\
        10 & Communication & com.whatsapp & 3.41 & 37.45 \\
        11 & Productivity & de.dhl.paket & 5.12 & 35.34 \\
        12 & Communication & org.telegram.messenger & \multicolumn{2}{l}{\textit{Decode Err.}} \\
        13 & Entertainment & com.ss.android.ugc.tiktok.pro & 7.93 & 38.21 \\
        14 & Shopping & de.payback.client.android & \multicolumn{2}{l}{\textit{GPlay Err.}} \\
        15 & Productivity & pdf.pdfreader.viewer.editor.free & 3.85 & 33.13 \\
        16 & Medical & shop.shop\_apotheke.com.shopapotheke & 3.6 & 36.19 \\
        17 & Productivity & com.google.android.apps.bard & 2.88 & 32.86 \\
        18 & Business & jump.readpdf.com & \multicolumn{2}{l}{\textit{GPlay Err.}} \\
        19 & Finance & com.google.android.apps.walletnfcrel & 3.14 & 33.53 \\
        20 & Communication & com.snapchat.android & 3.9 & 38.73 \\
        21 & Tools & com.google.android.apps.authenticator2 & 4.58 & 38.26 \\
        22 & Art \& Design & com.canva.editor & 4.9 & 33.38 \\
        23 & Medical & fr.doctolib.www & 5.37 & 43.8 \\
        24 & Food \& Drink & de.edeka.genuss & 3.19 & 35.22 \\
        25 & Shopping & com.lidl.eci.lidlplus & 3.72 & 36.11 \\
        26 & Shopping & fr.vinted & 3.13 & 35.74 \\
        27 & Shopping & com.myklarnamobile & 3.15 & 40.76 \\
        28 & Tools & com.gamma.scan & 1.84 & 38.09 \\
        29 & Tools & app.jumpjumpvpn.jumpjumpvpn & 2.9 & 38.66 \\
        30 & Lifestyle & com.pinterest & 3.42 & 37.05 \\
        31 & Entertainment & com.newleaf.app.android.victor & 3.11 & 38.37 \\
        32 & Shopping & com.zzkko & 3.43 & 36.18 \\
        33 & Social & com.newsclapper.video & 4.84 & 40.6 \\
        34 & Finance & com.paypal.android.p2pmobile & 3.56 & 41.84 \\
        35 & Finance & com.starfinanz.mobile.android.pushtan & \multicolumn{2}{l}{\textit{Monkey Err.}} \\
        36 & Shopping & de.idealo.android & 4.57 & 38.0 \\
        37 & Shopping & de.fabrik19.globus & 3.86 & 35.44 \\
        38 & Productivity & com.teacapps.barcodescanner & \multicolumn{2}{l}{\textit{GPlay Err.}} \\
        39 & Education & com.duolingo & 3.29 & 37.66 \\
        40 & Tools & com.sec.android.easyMover & 3.57 & 38.11 \\
        41 & Shopping & com.ebay.kleinanzeigen & 5.78 & 39.08 \\
        42 & Communication & com.whatsapp.w4b & 3.81 & 34.76 \\
        43 & Food \& Drink & de.rewe.app.mobile & 2.9 & 37.11 \\
        44 & Entertainment & de.prosiebensat1digital.seventv & 4.84 & 37.91 \\
        45 & Finance & de.check24.check24 & 2.91 & 34.49 \\
        46 & Finance & com.starfinanz.smob.android.sfinanzstatus & 2.94 & 41.4 \\
        47 & Maps \& Navigation & com.ubercab & 3.99 & 38.97 \\
        48 & Productivity & ai.x.grok & \multicolumn{2}{l}{\textit{GPlay Err.}} \\
        49 & Productivity & com.nect.app.prod & 3.9 & 40.18 \\
        50 & Shopping & com.kaufland.Kaufland & 9.72 & 36.55 \\
        \bottomrule
    \end{tabular}
    \label{tab:appendix-rq2a-p1}
\end{table}

\begin{table}[h]
    \centering
    \caption{Full completeness results for dataset (bottom).}
    \begin{tabular}{l|l|l|l|l}
        \toprule
        \thead{\#} & \thead{Category} & \thead{App Package} & \thead{FT} & \thead{WD}\\
        \midrule
        51 & Entertainment & com.amazon.avod.thirdpartyclient & 3.53 & 38.48 \\
        52 & Shopping & com.valuephone.vpnetto & 5.34 & 40.72 \\
        53 & Lifestyle & de.rossmann.app.android & 4.28 & 37.64 \\
        54 & Shopping & com.amazon.mShop.android.shopping & 3.45 & 37.47 \\
        55 & Productivity & com.adobe.reader & 4.26 & 39.79 \\
        56 & Food \& Drink & com.app.tgtg & 4.06 & 36.64 \\
        57 & Tools & com.google.android.apps.translate & 3.88 & 37.52 \\
        58 & Shopping & com.alibaba.aliexpresshd & 4.37 & 38.81 \\
        59 & Business & com.microsoft.teams & 3.39 & 33.14 \\
        60 & Business & com.azure.authenticator & 3.94 & 37.05 \\
        61 & Shopping & de.dm.meindm.android & 5.84 & 36.75 \\
        62 & Social & com.facebook.katana & 5.23 & 40.77 \\
        63 & Productivity & com.pdfviewer.prodoc & 3.69 & 35.89 \\
        64 & Productivity & com.alldocumentexplor.ade & 4.53 & 34.78 \\
        65 & Tools & com.cleanertool.box & 3.22 & 33.93 \\
        66 & Social & com.instagram.barcelona & 4.72 & 40.46 \\
        67 & Communication & com.imo.android.imoim & 5.48 & 36.96 \\
        68 & Communication & de.telekom.android.customercenter & 2.54 & 41.0 \\
        69 & Education & com.solocode.anton & 15.68 & 34.6 \\
        70 & Entertainment & de.rtli.tvnow & 3.12 & 32.32 \\
        71 & Entertainment & com.netflix.mediaclient & \multicolumn{2}{l}{\textit{Decode Err.}} \\
        72 & Communication & com.facebook.orca & 4.47 & 44.32 \\
        73 & Communication & com.discord & 3.44 & 39.78 \\
        74 & Tools & com.scannerreader.qrcode.creatorfree & 3.31 & 31.81 \\
        75 & Health \& Fitness & de.aoksystems.amg & 3.2 & 34.59 \\
        76 & Shopping & com.ingka.ikea.app & 3.43 & 38.55 \\
        77 & Productivity & open.chat.gpt.aichat.bot.free.app & 2.77 & 31.91 \\
        78 & Business & reader.pdfreader.com & \multicolumn{2}{l}{\textit{GPlay Err.}} \\
        79 & Communication & org.mozilla.firefox & 3.37 & 44.98 \\
        80 & Business & us.zoom.videomeetings & 2.94 & 38.42 \\
        81 & Shopping & com.ebay.mobile & 4.56 & 37.01 \\
        82 & Shopping & com.nintendo.znej & 3.99 & 37.1 \\
        83 & Lifestyle & com.point4more.ios.tak.user.prod & \multicolumn{2}{l}{\textit{GPlay Err.}} \\
        84 & Travel \& Local & net.easypark.android & 3.99 & 37.53 \\
        85 & Shopping & com.bonial.kaufda & 4.92 & 34.59 \\
        86 & Communication & org.thoughtcrime.securesms & 2.77 & 37.17 \\
        87 & Shopping & at.helloagain.muellerde & 4.06 & 42.19 \\
        88 & Music \& Audio & com.spotify.music & 3.74 & 36.64 \\
        89 & Shopping & com.action.consumerapp & 6.55 & 38.37 \\
        90 & Entertainment & com.storymatrix.drama & \multicolumn{2}{l}{\textit{GPlay Err.}} \\
        91 & Shopping & com.media.markt & 4.43 & 41.54 \\
        92 & Communication & com.truecaller & 4.21 & 35.59 \\
        93 & Finance & de.deutschepost.postident & \multicolumn{2}{l}{\textit{GPlay Err.}} \\
        94 & Finance & de.fiduciagad.securego.vr & \multicolumn{2}{l}{\textit{Monkey Err.}} \\
        95 & Entertainment & com.dramawave.app & 3.08 & 40.8 \\
        96 & Entertainment & com.disney.disneyplus & 4.66 & 36.47 \\
        97 & Entertainment & tv.twitch.android.app & 2.92 & 37.5 \\
        98 & News \& Magazines & com.twitter.android & 3.66 & 38.84 \\
        99 & Business & drive.readpdf.com & \multicolumn{2}{l}{\textit{GPlay Err.}} \\
        100 & Music \& Audio & com.suno.android & \multicolumn{2}{l}{\textit{GPlay Err.}} \\
        \bottomrule
    \end{tabular}
    \label{tab:appendix-rq2a-p2}
\end{table}